\documentstyle[12pt,epsf]{article}

\begin{document}

\baselineskip=20pt
\newcommand{\pb}{\mbox{e}^{\hat{\Phi}}}
\renewcommand{\theequation}{\arabic{section}.\arabic{equation}}

\ \vspace{2cm}

\begin{center}
   {\Large \bf 
           Intermediate coherent-phase(PB) states of radiation 
           fields and their nonclassical properties}\\ \ \\

   {\large \bf 
           Yongzheng Zhang$^*$, Hongchen Fu$^\dagger $ 
           \footnote{E-mail: h.fu@open.ac.uk} and Allan I. Solomon$^\dagger $ 
           \footnote{E-mail: a.i.solomon@open.ac.uk}}\\
   {\it    * Department of Mathematics, Northeast Normal 
           University, \\ Changchun 130024, P.R.China}\\

   {\it    $\dagger$ Faculty of Mathematics and Computing, 
           The Open University, \\ Milton Keynes,
           MK7 6AA, U.\,K.}
\end{center}

\vspace{4cm}

\begin{abstract}
Intermediate states interpolating coherent states and 
Pegg-Barnett phase states are investigated using 
the ladder operator approach. These states reduce 
to coherent and Pegg-Barnett phase states in two different 
limits. Statistical and squeezing properties are studied in
detail. 
\end{abstract}

\newpage

%%%%%%%%%%%%%%%%%%%%%%%%%%%%%%%%%%
\section{Introduction and Motivation}
\setcounter{equation}{0}
%%%%%%%%%%%%%%%%%%%%%%%%%%%%%%%%%%

Since Stoler {\it et al}\/ introduced the binomial states (BS)
in 1985 \cite{stol}, the so-called intermediate states interpolating
two fundamental states of radiation fields have
attracted much interests in quantum optics [1-13]. The BS is
defined as a linear superposition of number states in an
$(M+1)$-dimensional subspace
\begin{equation}
  |\eta,\,M\rangle=\sum_{n=0}^{M}\left[\beta_n^M(\eta)
  \right]^{\frac{1}{2}}|n\rangle,   \label{bs}
\end{equation}
where $\eta$ is a real parameter satisfying $0<\eta<1$, and
\begin{equation}
  \beta_n^M(\eta)={M \choose n}\eta^n (1-\eta)^{M-n}
  \label{bsdist}
\end{equation}
is the binomial distribution with probability $\eta$. In the
limits $\eta \to 1$ and  $\eta \to 0$, BS reduce to number
states $|1,\,M\rangle=|M\rangle$ and $|0,\,M\rangle=|0\rangle$,
respectively. In a different limit of $M\to \infty, \ \eta\to 0$
with $\eta M =\alpha^2$ fixed ($\alpha$ real constant)
$|\eta,\,M\rangle$ reduce to the  coherent states
with {\it real} amplitude $\alpha$. In this sense, BS are
the {\it intermediate number-coherent states}. The notion of
BS was also generalized to the multinomial \cite{fus7} and
negative multinomial states \cite{fus7,fus6}, hypergeometric
states \cite{fus5}, P\'{o}lya states \cite{fus8}, intermediate
number-squeezed states \cite{fus4,bas1} and the number-phase
states \cite{bas2}, as well as its $q$-deformation \cite{fan1}.

In a previous paper \cite{fus4} one of the authors
presented a ladder operator formalism of BS, namely, 
BS satisfy the following eigenvalue equation
\begin{equation}
    \left(\sqrt{\eta}N+\sqrt{1-\eta}J^+_M \right)|M,\eta\rangle
    =\sqrt{\eta}M |M,\eta\rangle,  \label{hh}
\end{equation}
where $J^+_M=\sqrt{M-N}\,a\, $ is the raising operator of su(2)
 via
its Holstein-Primakoff realization. We also proposed the
generalized BS by replacing $J^+_M$ with a linear combinatio
of $J^+_M$ and $J^-_M\equiv (J^+_M)^{\dagger}$ which are the
intermediate number-squeezed states. From this approach we
learn that (1) the parameter $\eta$ plays the role of controlling
two different limits and (2) the limit to coherent states is
essentially the contraction of Lie algebra su(2) to the oscillator
algebra: $\sqrt{\eta}J_M^+ \to \alpha a$
in the limit $\eta\to 0$ and $M\to \infty$ with $\eta M=\alpha^2$.
So in the ladder operator approach of an intermediate state we can us
su(2) generators to control the coherent state limit.

In this letter we shall pay our attention to the
intermediate states between coherent states and the
Pegg-Barnett (PB) phase states, which, to our knowledge,
are not considered in the literature. We shall generalize
the ladder operator approach of BS to these {\it intermediate
coherent-phase(PB) states} (ICPS). Above discussion on BS
suggest us proposing the following eigenvalue equation
\begin{equation}
  \left( \sqrt{\eta}\pb+\sqrt{1-\eta} J_M^+
  \right)|M,\eta,\rho\rangle
  =\rho |M,\eta,\rho\rangle.     \label{equation}
\end{equation}
Here $0 <\eta <1$ is real as in the BS case, and 
$\rho$ is eigenvalue to be determined. The
operator $\pb$ is the exponential PB phase operator
defined by \cite{pbpo}
\begin{equation}
    \pb|\theta_m\rangle=e^{\theta_m}|\theta_m\rangle
\end{equation}
on the PB phase state
\begin{equation}
    |\theta_m\rangle=\frac{1}{\sqrt{M+1}}\sum_{n=0}^M
                     \exp{(in\theta_m)}|n\rangle,\ \ \ \
    \theta_m=\frac{2\pi m}{M+1}+\theta_0, \label{thetam}
\end{equation}
where $\theta_0$ is a real constant.

We shall solve the equation (\ref{equation}) in next section an
then discuss its limits to coherent and PB phase states in Sec.3.
The photon statistics and the squeezing properties are investigated
in detail in Sec.4. Sec.5 is a concluding remark. We note that
these states are shown to be finite superposition of Fock states and
in principle can be experimentally fabricated,
as reported recently in \cite{jdsa}

%%%%%%%%%%%%%%%%%%%%%%%%%%%%%%%%%%%%%%%%%%%%%%%%%%%%%%%%%%
\section{Intermediate coherent-phase(PB) states}
\setcounter{equation}{0}
%%%%%%%%%%%%%%%%%%%%%%%%%%%%%%%%%%%%%%%%%%%%%%%%%%%%%%%%%%
 
Equation (\ref{equation}) is an eigenvalue equation of an
$(M+1)\times(M+1)$ matrix, so it has $M+1$ eigenvalues and
corresponding eigenstates. To solve it, we expand the state
$|M,\eta,\rho\rangle$ in terms of the number state
\begin{equation}
   |M,\eta,\rho\rangle=\sum_{n=0}^M C_n |n\rangle.
   \label{expand}
\end{equation}
Inserting (\ref{expand}) into (\ref{equation}) and using
the following relations \cite{fus2}
\begin{equation}
   \pb|n\rangle=|n-1\rangle \ \ \ \ (n\neq 0),\ \ \ \
   \pb|0\rangle=e^{i(M+1)\theta_0}|M\rangle,
\end{equation}
we obtain the following equations
\begin{eqnarray}
&&  \sqrt{\eta}C_0 e^{i(M+1)\theta_0}=\rho C_M, \label{con}\\
&&  \left(\sqrt{1-\eta}\sqrt{n(M-n+1)}+\sqrt{\eta}\right)
    C_n=\rho C_{n-1} \ \ \ \ (n=1,\cdots,M).  \label{coef}
\end{eqnarray}
From (\ref{coef}) we have
\begin{equation}
    C_n=\frac{\rho^n}{F(n)!}C_0  \ \ \ \ (n=1,\cdots,M),
    \label{cn}
\end{equation}
where
\begin{eqnarray}
&&    F(n)=\sqrt{1-\eta}\sqrt{n(M-n+1)}+\sqrt{\eta},\\
&&    F(n)!=F(n)F(n-1)\cdots F(1),\ \ \  F(0)!\equiv 1. 
\end{eqnarray}
Relation (\ref{cn}) with $n=M$ must be consistent with the
condition (\ref{con}), namely,
\begin{equation}
    \sqrt{\eta}C_0 e^{i(M+1)\theta_0}=
    \frac{\rho^{M+1}}{F(M)!}C_0,
\end{equation}
which leads to $M+1$ distinct eigenvalues ($C_0\neq 0$)
\begin{equation}
    \rho_m=\left(\sqrt{\eta}F(M)!\right)^{\frac{1}{M+1}}
    e^{i\theta_m}, \ \ \ \
    0\leq m\leq M,       \label{eigenvalue}
\end{equation}
where $\theta_m$ is the same as in Eq.(\ref{thetam}).
The normalization constant $C_0$ can  be easily determined as
\begin{equation}
    C_0=\left[ \sum_{n=0}^M \left(\frac{[
    \sqrt{\eta}F(M)!]^{\frac{n}{M+1}}}{F(n)!}
    \right)^2
    \right]^{-\frac{1}{2}}.
\end{equation}
Substituting (\ref{eigenvalue}) into (\ref{expand}) we finally
find the ICPS (we write 
$|M,\eta,\rho_m\rangle \equiv |M,\eta,\theta_m\rangle $
\begin{eqnarray}
&&    |M,\eta,\theta_m\rangle=\sum_{n=0}^M D_n^M(\eta
      e^{i\theta_m n} |n\rangle , \\
&&    D_n^M(\eta)=\left(\sum_{n=0}^M \left(\frac{[
      \sqrt{\eta}F(M)!]^{\frac{n}{M+1}}}{F(n)!}\right)^2
      \right)^{-\frac{1}{2}}
      \frac{\left(
      \sqrt{\eta}F(M)!\right)^{\frac{n}{M+1}}}{F(n)!}
\end{eqnarray}
Here we have written  $ e^{i\theta_m n} $ separately
for convenience in later use.

It is interesting that using the identity method in \cite{fus2}
these states can also be written as the form of a {\it
displacement operator} acting on the vacuum state
\begin{equation}
    |M,\eta,\theta_m\rangle=C_0\exp_M
      \left(\frac{[\sqrt{\eta} F(M)!]^{\frac{1}{M+1}}
      \sqrt{N}}{F(N)}\, a^{\dagger}\right)|0\rangle,
\end{equation}
where $\exp_M(x)=\sum_{n=0}^{M}x^n/n!$ is the
{\it finite} exponential function.

The parameter $\theta_0$ ($0\leq \theta <2\pi$) has clear
physical meaning: it reflects the time development of 
ICPS. This can be seen from $e^{-iHt}|M,\eta,\theta_m\rangle
=|M,\eta,\theta_m-\omega t\rangle$, where $H=\omega (N+1/2)$
is the Hamiltonian of the single mode radiation field. In
next section we shall see that in the coherent limit,
$\theta_0$ do gives the imaginary part of amplitude of
limiting coherent states which reflects the time evolution
of coherent states.

%%%%%%%%%%%%%%%%%%%%%%%%%%%%%%%%%%%%%%%%%%%%%%%%%%%
\section{Limits to PB phase states and coherent states}
\setcounter{equation}{0}
%%%%%%%%%%%%%%%%%%%%%%%%%%%%%%%%%%%%%%%%%%%%%%%%%%%%
 
We first consider the limit $\eta\to 1$. It is easy to see that
\begin{equation}
    \frac{\left[
    \sqrt{\eta}F(M)!\right]^{\frac{n}{M+1}}}{F(n)!}\to 1,\ \ \ 
    C_0\to \left(\sum_{n=0}^M 1\right)^{-1/2}=
    \frac{1}{\sqrt{M+1}}
\end{equation}
So
\begin{equation}
    |M,\eta,\theta_m\rangle \to
    \frac{1}{\sqrt{M+1}}\sum_{n=0}^M e^{i\theta_m n}|n\rangle
    \equiv |\theta_m\rangle.
\end{equation}
We arrive at the PB phase states.
 
In a different limit: $M\to\infty$, $\eta\to 0$ keeping
$\eta(M/\alpha)^{M+1}=\beta$ a finite constant ($\alpha$ is a
real constant), we will get the coherent states.
In this limit, $ F(n)\to \sqrt{nM} $,  $ F(M)!\to M! $
and     
\begin{equation}
    (M+1)^{\frac{1}{M+1}}\sim 1,\ \ \ 
    M\sim M+1\ \ \ \ \mbox{for $M\to\infty$}.
\end{equation}
We then have
\begin{equation}
    \frac{[\sqrt{\eta}F(M)!\,]^{\frac{n}{M+1}}}{F(n)!}=
    \frac{1}{\sqrt{n!}}
    \left[\lim_{\scriptsize\begin{array}{c}M\to\infty\\
    \eta\to 0\end{array}}
    \left(\sqrt{\eta M^{M+1}}\right)^{\frac{1}{M+1}}
    \lim_{M+1\to\infty}
    \frac{[(M+1)!]^{\frac{1}{M+1}}}{M+1}\right]^n.
\end{equation}
By making use of the following limit formula
\begin{equation}
   \lim_{M+1\to\infty}\frac{[(M+1)!]^{\frac{1}{M+1}}}{M+1}=
   \frac{2}{e},
\end{equation}
we have
\begin{equation}
   \frac{[\sqrt{\eta}F(M)!]^{\frac{n}{M+1}}}{F(n)!}\to
   \frac{1}{\sqrt{n!}}  \left(\frac{2}{e}\alpha\right)^n.
   \label{limit1}
\end{equation}
In this limit, $C_0$ and $\theta_m$ reduce to
\begin{equation}
   C_0\to \exp{\left(-\frac{2\alpha^2}{e^2}\right)}, \ \ \ 
   \theta_m\to \theta_0. 
   \label{limit2}
\end{equation}
From (\ref{limit1}) and (\ref{limit2}) 
we obtain the coherent state limit
\begin{equation}
   |M,\eta,\theta_m\rangle \to \left|\frac{2\alpha}{e}
   e^{i\theta_0}\right\rangle
   \equiv \exp{\left(-\frac{2\alpha^2}{e^2}\right)} 
   \sum_{n=0}^{\infty}\frac{\left(
   \frac{2\alpha}{e}e^{i\theta_0}\right)^n}{\sqrt{n!}}
   |n\rangle.
\end{equation}
We note that $M+1$ different ICPS reduce to the same
coherent state due to $\theta_m\to\theta_0$ for all
$m$.
 
We also remark that, similarly to the BS, the intermediate
coherent-(PB)phase states degenerate to the vacuum state in
the limit $\eta\to 0$.
 
%%%%%%%%%%%%%%%%%%%%%%%%%%%%%%
\section{Nonclassical properties}
\setcounter{equation}{0}
%%%%%%%%%%%%%%%%%%%%%%%%%%%%%%
 
\subsection{Photon statistics}
 
Mandel's $Q$-factor characterizing sub(super)-Poissonian
distribution is obtained as
\begin{equation}
  Q(M,\eta)=\frac{\langle\Delta N^2\rangle}{\langle N \rangle}-1
  =\frac{\sum_{n=0}^M [D^M_n(\eta)]^2 n^2 -
  \left[\sum_{n=0}^M [D^M_n(\eta)]^2 n\right]^2}
  {\sum_{n=0}^M [D^M_n(\eta)]^2 n}-1.
\end{equation}
If $Q<0 \ (>0)$, the field is of sub(super)-Poissonian.
$Q=0$ corresponds to the Poissonian statistics.  We note that
$Q(M,\eta)$ is independent of the parameter $\theta_m$ and
therefore it reflects the photon statistics of all $M+1$ state
$|M,\eta,\theta_m\rangle$.  
 
Fig.1. is a plot of $Q(M,\eta)$ as a function of $\eta$ for
different values of $M$ $(1,2,\cdots,7)$. We find that the
field on ICPS is of sub-Poissonian in the case $M=1$ except for
the end point $\eta=0$. For the cases $M=2,3$, the field becomes
super-Poissonian first from the Poissonian statistics at
$\eta=0$, and then the sub-Poissonian. The range of the
sub-Poissonian statistics for $M=2$ is wider than that for
$M=3$. When $M=4$, the fields are of super-Poissonian except
for two end points $\eta=0,1$, which correspond Poissonian. Finally,
if $M>4$, the fields are super-Poissonian except for the
starting point $\eta=0$. We here note that the $Q$-factor of
the PB phase states is $Q(M,0)=(M-4)/6$, which correspond to
the right ends of the $Q$ factor.
 
%%%%%%%%%%%%%%%%%%%%%%%%%%%%%%%%
\subsection{Squeezing effect}
%%%%%%%%%%%%%%%%%%%%%%%%%%%%%%%%
 
Define the coordinate $x$ and the momentum $p$ as
\begin{equation}
       x=\frac{1}{\sqrt{2}}(a^\dagger+a),\ \ \ 
       p=\frac{i}{\sqrt{2}}(a^\dagger-a).
\end{equation}
Then their variances 
$ (\Delta x)^2\equiv \langle x^2\rangle-\langle x\rangle^2 $ 
and
$ (\Delta p)^2\equiv \langle p^2\rangle-\langle p\rangle^2 $ 
are obtained as
\begin{eqnarray}
  (\Delta x)^2 &=&
       \frac{1}{2}+\sum_{n=0}^M n D^M_n(\eta)^2 +
       \cos(2\theta_m)\sum_{n=0}^{M-2}\sqrt{(n+1)(n+2)}
       D^M_n(\eta) D^M_{n+2}(\eta) \nonumber \\ 
       && - 2\left(\cos(\theta_m)\sum_{n=0}^{M-1}\sqrt{n+1}
       D^M_n(\eta) D^M_{n+1}(\eta)\right)^2,   \nonumber\\
  (\Delta p)^2 &=&
       \frac{1}{2}+\sum_{n=0}^M n D^M_n(\eta)^2 -
       \cos(2\theta_m)\sum_{n=0}^{M-2}\sqrt{(n+1)(n+2)}
       D^M_n(\eta) D^M_{n+2}(\eta) \nonumber \\ 
       && + 2\left(\sin(\theta_m)\sum_{n=0}^{M-1}\sqrt{n+1}
       D^M_n(\eta) D^M_{n+1}(\eta)\right)^2.
\end{eqnarray}
If $(\Delta x)^2 <1/2 $
(or $(\Delta p^2) <1/2$), we say the quadrature $x$ 
(or $p$) is squeezed.
 
We first note that $(\Delta x)^2$ and
$(\Delta p)^2$ are related with each other
by the following relation
\begin{equation}
      (\Delta x)^2_{\theta_m}=
      (\Delta p)^2_{\theta_m\pm\pi/2}.
\end{equation}
So hereafter we only consider the quadrature
$(\Delta x)^2$. Then it is obvious that
$(\Delta x)^2$ is a $\pi$-periodic
function of $\theta_m$ and it is symmetric
with respect to $\theta_m=\pi/2$.
 
Figures 2 shows how $(\Delta x)^2 $
depends on parameters $M$, $\eta$ and 
$\theta_m$, respectively. From these plots we find that
 
1. When $\theta_m=\pi/2$. In this case the quadrature $x$ is not
squeezed at the point $\eta=0$, which corresponds to the
vacuum state. Then, with
the increase of $\eta$, it becomes squeezed drastically until
the maximum of squeezing (minimum of $(\Delta x)^2$) 
is reached.  By further increasing $\eta$,  the
squeezing becomes weaker and weaker until it
disappears for a large enough $\eta_0$. 
The squeezing range $0<\eta<\eta_0$
depends on $M$: the larger $M$, the wider the 
squeezing range and the smaller $(\Delta x)^2$. 
 
2. Dependence on $\theta_m$.  Since $(\Delta x)^2$ 
is symmetric with respect to $\theta_m=\pi/2$, so 
we only plot $\theta_m\leq \pi/2$ part in Fig.2.  
We see that, with the decreas
(or increase) of $\theta_m$ form $\pi/2$, the squeezing
becomes weaker and weaker and the squeezing range 
$0<\eta<\eta_0$ for a fixed $\theta_m$ becomes 
narrower and narrower, until squeezing 
disappears for small (or large) enough $\theta_m$. 
 
%%%%%%%%%%%%%%%%%%%%%
\section{Conclusion}
%%%%%%%%%%%%%%%%%%%%%
 
In this paper we have introduced the intermediate
coherent-phase(PB) states by ladder
operator approach and investigated
their nonclassical properties. As the intermediate states,
these states interpolate between the coherent states and
the PB phase states and reduce to them in two different
limits. They also exhibit strong nonclassical properties
such as sub-Poissonian statistics and squeezing effect in
considerable ranges of parameters involved. 
 
Finally, we point out
that, as a finite superposition of Fock states, these
states in principle can be experimentally fabricated,
as reported recently \cite{jdsa}.

%%%%%%%%%%%%%%%%%%%%%%%%%%%
\section*{Acknowledgments}
%%%%%%%%%%%%%%%%%%%%%%%%%%
This work is supported in part by the National Natural
Science Foundation of China through Northeast Normal
University (19875008).
   
%%%%%%%%%%%%%%%%%%%%%%%%%%%%%%%%%%%%%%%%%%%%%%%%%%%

%%%%%%%%%%%%%%%%%%%%%%%%%%%%%%%%%%%%%%%%%%%%%
\newpage %%%%%%%%%%%%%% FIGURES %%%%%%%%%%%%%
%%%%%%%%%%%%%%%%%%%%%%%%%%%%%%%%%%%%%%%%%%%%%

\begin{figure}
\centerline{\epsfxsize=9cm
\epsfbox{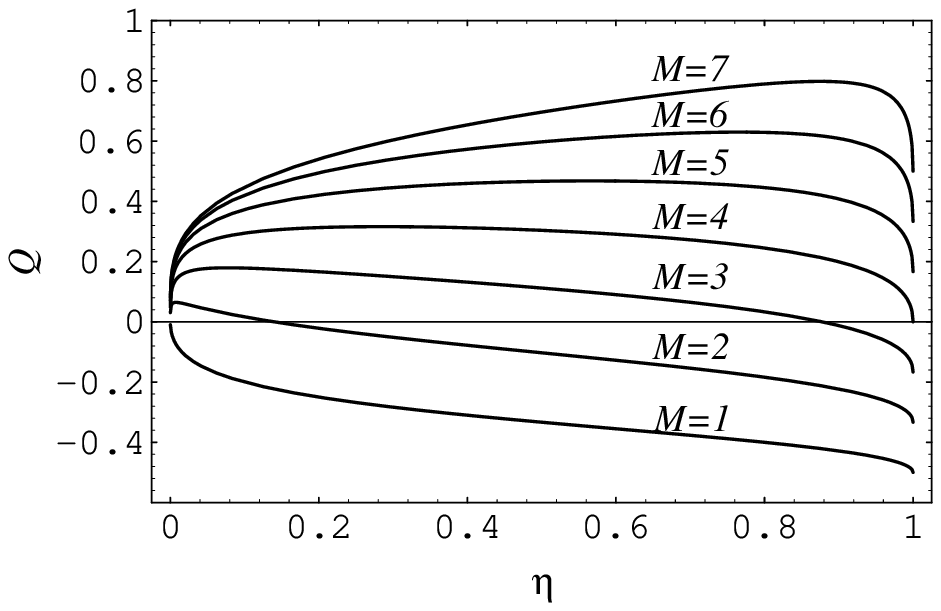}}
\caption{The Mandel's $Q$-factor $Q(M,\eta)$ as a funtion
            of $\eta$, for different $M=1,\cdots,7$.}
\end{figure}

\begin{figure}
\centerline{\epsfxsize=8cm
\epsfbox{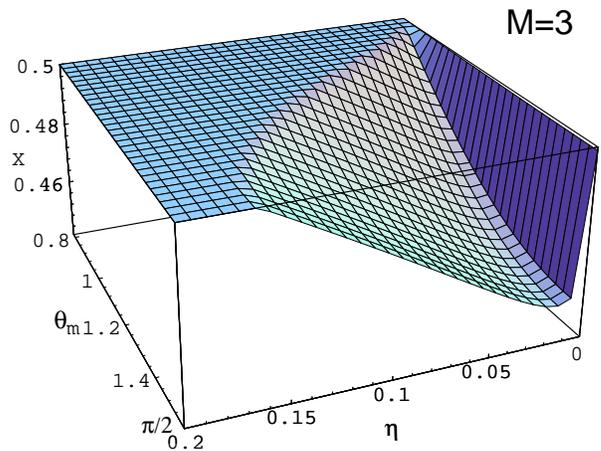}} \vspace{1cm}
\centerline{\epsfxsize=8cm
\epsfbox{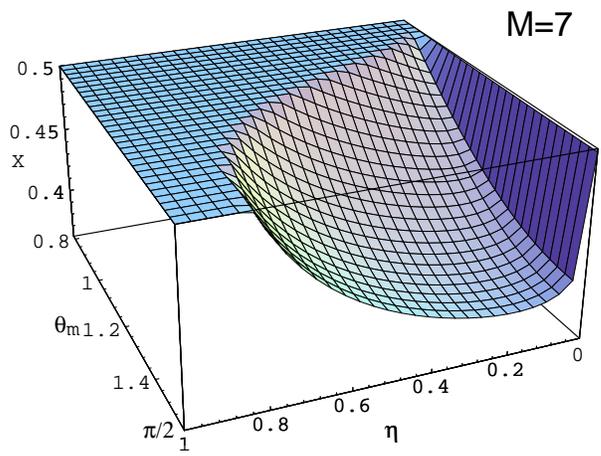}}
\caption{Variance $\langle \Delta x^2\rangle\equiv X$ as a function
of $\eta$ and $\theta_m$ for $M=3,\,7$.}
\end{figure}

\end{document}